# Phase-Only Beam Shaping for Transmitting Array Antennas in Radar Applications

Lior Maman, *Student Member, IEEE*, Shlomo Zach, and Amir Boag, *Fellow, IEEE*

*Abstract*—Two beam broadening methods for active electronically scanned array (AESA) antennas with uniform amplitude excitation are proposed and compared: phase tapering optimization (PTO) and a novel time-varying phase tapering (TPT). The PTO is a simple and efficient approach assuming continuous polynomial phase distribution and requiring optimization of only few parameters. The TPT is valid mainly for radar applications, taking advantage of the fact that radars typically transmit pulse trains for coherent integration. By varying the array elements' phases from pulse to pulse, the TPT achieves effective amplitude tapering, thus providing a method of beam shaping, occasionally with a simple analytic form. The TPT also makes it possible to produce beam shaping with very low side lobe levels in comparison to the PTO. As a preliminary step, the dimensionality of the radiation pattern characterization for all scan directions is reduced from five to only two variables. This is crucial for efficient optimization of the radiation pattern which needs to be evaluated over a judiciously specified two-dimensional domain.

## I. Introduction

Radiation pattern synthesis and, in particular, beam shaping are essential for the design and performance optimization of active electronically scanned array (AESA) antennas. The beam shaping, in most cases, involves beam broadening, reducing the side lobe level (SLL), or generating a cosecant squared pattern [1-3]. This paper will focus on beam broadening and reducing the SLL in the transmitting mode, which are important in many radar applications.

An AESA antenna typically comprises a large number of relatively low-power radiating elements, making it geometrically large. The transmitting power of each element is limited due to reliability, cost, and thermal considerations. The large electrical size of the transmitting antenna tends to produce a narrow illuminating beam, thus limiting the scan sector and revisit time [4]. This problem is especially acute for ground surveillance radars [5] which are designed to rapidly scan a large search sector and deal with multiple targets in many directions in parallel. For these radars, extreme beam broadening is usually required, while for multi-beam mode [6], moderate beam broadening is usually sufficient. The main benefits of the beam broadening include providing more coverage area for search and detection, allowing faster revisit time, and improving the time on target (TOT), which is crucial for Doppler resolution [7].

Typically, in conjunction with the beam broadening, low SLL is also required. Low SLL is crucial in designing radar systems since it reduces false alarm detections, prevents saturation from clutter, and improves the dynamic range of the receivers [8-9]. SLL can be significantly reduced with amplitude tapering [10], however, most modern AESA antennas are based on solid-state amplifiers that, for the best efficiency, have to operate in the saturation mode. In other words, efficiency and thermal issues preclude conventional amplitude tapering for transmitted beam shaping. Thus, beam broadening and reducing SLL must be achieved with uniform amplitude using phase control only.

Various approaches to this non-trivial task have been proposed in the literature. An interesting idea proposed in [11] allows producing effective amplitude tapering by using opposite phases, i.e. conjugate excitation coefficients, in adjacent elements. Notably, this approach is applicable only for small inter-element spacings, on the order of quarter wavelength, as demonstrated in the article. This limitation might be incompatible with element sizes and, furthermore, may reduce gain and influence mutual coupling between the elements. On the other hand, the concept of such spatially alternating phases can be replaced by time modulation, particularly in radar implementations, to mitigate the aforementioned drawbacks, a topic we will explore in further detail later.

The gradient search algorithm optimization approach to phase tapering was proposed in [12]. In [12] and [13], beam broadening was achieved with stochastic gradient descent and genetic algorithm (GA), respectively. These methods are highly time-consuming since they optimize the phase shift of every element, and there is no a priori information on the shape of the phase distribution. Since beam broadening and phase tapering are mainly used for radars transmitting pulse trains, one can consider time-dependent modulation. The time modulated arrays (TMA) have been proposed and studied in [14-30], where the radiation pattern can be synthesized by controlling the transmission time of each element. Unfortunately, precisely controlling the transmission time in large arrays may be expensive.

This paper, building upon our preliminary work [31], focuses on two methods for beam broadening and reducing the SLL for all scan directions. The first one is the phase tapering optimization (PTO) and is considered mainly as a more conventional reference solution. Here, the elements' phases are tapered using a polynomial distribution determined by optimizing only few parameters, somewhat similar to [32-33]. Unlike stochastic optimization, this approach results in a continuous phase distribution over the radiating aperture. The second approach we propose and study is the novel time-varying phase tapering (TPT), which takes advantage of the typical radars that transmit pulse trains. We show that by coherent integration of alternating phase pulse returns, one can create effective amplitude tapering, even for moving targets. By providing this equivalence between phase and amplitude tapering, one can realize almost any desired pattern with very low SLL and, in some cases, in closed analytical form. For both methods, we a priori reduce the radiation pattern characterization problem from five variables to only two, thus making the computational task feasible in terms of computation time and memory.

This paper is organized as follows: Section II describes the settings, assumptions, and complexity, of the beam broadening problem for AESA. Section III presents an efficient array factor (AF) evaluation method by reducing the dimensionality of scanning AF from five variables, namely, elevation and azimuth angles of the observation and scanning directions, and frequency (wave number) to only two variables and obtaining the new domain. In Sections IV and V, we present the PTO and TPT algorithms, respectively. Section VI presents the numerical results of both algorithms for the beam broadening problem, demonstrating the advantages of the TPT in comparison with the more conventional phase tapering optimization.

## II. Problem Formulation

Consider the design of a planar AESA comprising identical T/R modules destined primarily for radar applications. In such configurations, it is often prescribed that only uniform amplitude aperture distributions are feasible in the transmitting mode due to efficiency and thermal management considerations. To that end, we focus on scanning and beam shaping in the transmitting mode using phase-only control of the element excitation.

Specifically, the phased array element lattice is assumed to lay in the $yz$-plane as depicted in Fig. 1. Here, $\alpha$ and $\varphi$ are the elevation and azimuth angles, respectively. The array is radiating into the half space $x > 0$, so the observation sector is defined as $-90° \leq \varphi \leq 90°$, $-90° \leq \alpha \leq 90°$. The main lobe (ML) direction $(\alpha_0, \varphi_0)$ is steered to cover a certain scan sector of interest defined by $\alpha_{0_{\min}} \leq \alpha_0 \leq \alpha_{0_{\max}}$ and $\varphi_{0_{\min}} \leq \varphi_0 \leq \varphi_{0_{\max}}$, where $\alpha_{0_{\min}}$, $\alpha_{0_{\max}}$ and $\varphi_{0_{\min}}$, $\varphi_{0_{\max}}$ are the minimum, maximum elevation and azimuth scan angles, respectively. In wideband applications, the radiation properties need to be characterized over a prescribed frequency band $f_{\min} \leq f \leq f_{\max}$.

In this paper, for the sake of simplicity, we will focus only on the array factor (AF), which means that we effectively assume that the element pattern is almost isotropic. We also assume that the mutual coupling between the elements is negligible or can be compensated for by a proper excitation. The AF can then be computed as:

$$\text{AF}(k, \alpha, \varphi, \alpha_0, \varphi_0) = \sum_{n=1}^{N_e} I_n e^{jk r_n \cdot (\hat{r} - \hat{r}_0)} \quad (1)$$

where $k$ is the wavenumber, $\hat{r} = \cos\alpha \cos\varphi \hat{x} + \cos\alpha \sin\varphi \hat{y} + \sin\alpha \hat{z}$ is a unit vector in the direction of observation, $\hat{r}_0 = \cos\alpha_0 \cos\varphi_0 \hat{x} + \cos\alpha_0 \sin\varphi_0 \hat{y} + \sin\alpha_0 \hat{z}$ is the main beam direction, while $I_n$ and $r_n = y_n \hat{y} + z_n \hat{z}$ are the excitation coefficient (excluding linear phase $e^{-jk r_n \cdot \hat{r}_0}$ used for scanning) and location of the $n$th element, respectively. From (1), we can infer that the AF depends on five variables, namely, frequency via $k$, observation direction $(\alpha, \varphi)$, and main beam direction $(\alpha_0, \varphi_0)$.

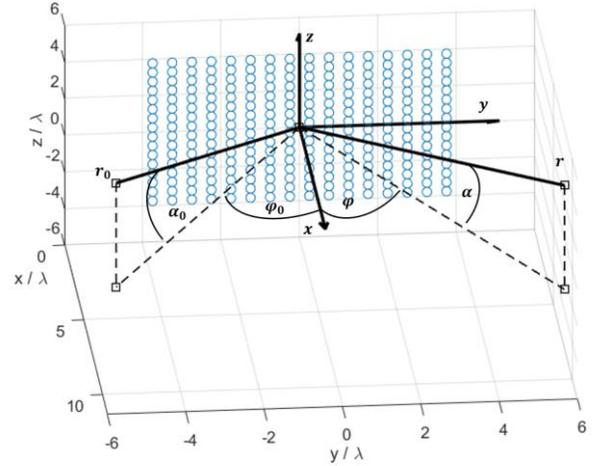

Fig. 1. A planar phased array antenna comprising $16 \times 16$ elements laying in the $yz$-plane and radiating into the $x > 0$ half space.

Beam shaping in the transmitting mode entails adjusting the antenna pattern to cover a certain search sector efficiently. In this work, we will focus on a typical example of beam shaping, namely, beam broadening. Beam broadening is required when a radar antenna beam is too narrow in the search mode. This means that for a reasonable revisit time, the TOT would be too short, which spoils the Doppler resolution. Thus, beam broadening is necessary for large AESA antennas to allow high transmitting power while sustaining good Doppler resolution.

In the beam broadening problem, we aim to broaden the transmitted beam while simultaneously maximizing the energy in the main lobe (ML) region, minimizing the energy outside the ML, and minimizing the ripple within the ML to illuminate a specific search sector with a uniform radiated power. Fig. 2 presents an example of the ML and ripple defined in the beam broadening problem.

As already stated, only uniform amplitude weighting is possible in the transmitting mode, which means that the constraint on $I_n$ is $|I_n| = I_0$, where $I_0$ is a constant. This constraint makes the beam broadening problem challenging. To the best of the authors' knowledge, the problem has no analytic solution yet, and in most cases, the solution is obtained by optimization.

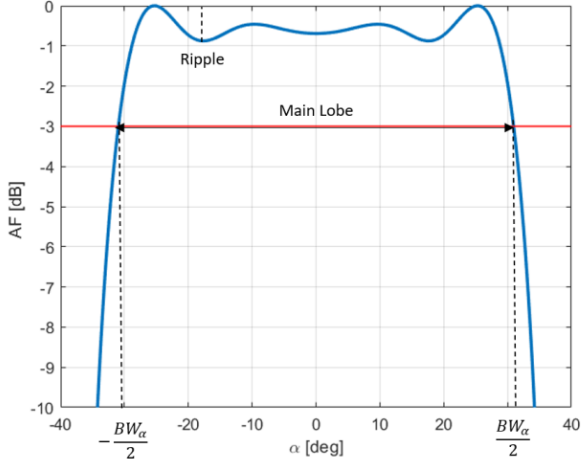

Fig. 2. Example of radiation pattern in elevation cut for beam broadening problem.

Unfortunately, there are two main difficulties in applying optimization to the scanning phased array problem. First, it is computationally very heavy since the AF needs to be evaluated in a five-dimensional space (AF depends on $k, \alpha, \varphi, \alpha_0$ and $\varphi_0$). Second, independent of the type of cost function, the optimization problem is not convex, thus tending to converge to local minima. In the following sections, we provide methods to overcome these difficulties.

## III. EFFICIENT ARRAY FACTOR EVALUATION

In this Section, we reduce the AF dimensionality from five to only two variables within a bounded domain which is crucial for an efficient optimization. Such transformation enables displaying the AF for all scanning directions and frequencies on a single 2-D domain to be determined below.

The AF was expressed in (1) as a function of 5 variables. Instead, we can express it via the observation direction wavevector, $\boldsymbol{k} = k\hat{\boldsymbol{r}}$, whose $yz$-components are $k_y = k\cos\alpha \sin\varphi$ and $k_z = k\sin\alpha$, and the main beam direction wavevector, $\boldsymbol{k}_0 = k\hat{\boldsymbol{r}}_0$, where $k_{y_0} = k\cos\alpha_0 \sin\varphi_0$ and $k_{z_0} = k\sin\alpha_0$. Combining the frequency (wavenumber) with the observation and scanning directions, we define a new vector $\boldsymbol{u} = k(\hat{\boldsymbol{r}} - \hat{\boldsymbol{r}}_0) = \boldsymbol{k} - \boldsymbol{k}_0$. This definition combines the conventional direction sines [34] and frequency. Since, $\boldsymbol{r}_n$ is in the $yz$-plane, the $x$-component of $\boldsymbol{u}$ does not affect the AF, implying that it depends only on $\boldsymbol{u}_T = \boldsymbol{k}_T - \boldsymbol{k}_{0_T} = (u_y, u_z)$ where $u_y = k_y - k_{y_0}$ and $u_z = k_z - k_{z_0}$. This allows us to express the AF in terms of these two variables only:

$$\text{AF}(u_y, u_z) = \sum_{n=1}^{N_e} I_n e^{j\boldsymbol{r}_n \cdot \boldsymbol{u}_T} \qquad (2)$$

where, obviously, $\boldsymbol{r}_n \cdot \boldsymbol{u}_T = y_n u_y + z_n u_z$.

Now, we have to find the relevant support of the AF in the $\boldsymbol{u}_T$ plane. While this support can be found by a brute-force scanning of the five-dimensional space of the original variables, here we outline an analytic technique for determining its outer boundary (detailed description of this procedure can be found in Appendix A). Since the wavenumber produces merely scaling of $\boldsymbol{u}_T$ the boundary should be determined for $k$ corresponding to the highest frequency of interest. An AF support boundary for $\varphi_{0\min} = -25°$, $\varphi_{0\max} = 25°$, $\alpha_{0\min} = -25°$ and $\alpha_{0\max} = 25°$ is shown in the $\boldsymbol{u}_T$-plane normalized to $k$ in Fig. 3. The boundary of $\boldsymbol{u}_T = \boldsymbol{k}_T - \boldsymbol{k}_{0_T}$ is obtained by considering the observation and scanning domains described by the $\boldsymbol{k}_T$ and $\boldsymbol{k}_{0_T}$ boundaries, respectively. In Fig. 3, the observation domain is bounded by a unit circle (dashed line), while the scanning one appears as a slightly "distorted" rectangle (dotted line). Without loss of generality, we analyze the problem for the first quadrant, i.e., positive values of $u_y$ and $u_z$. The points A, B, C, $A_0$, $B_0$ and $C_0$ in Fig. 3 represent the key points determining three different segments of the $\boldsymbol{u}_T$-boundary. Segment I (shown in Fig. 3 as a thick black line) starts at point $P = \overline{OA} - \overline{OA_0}$ and ends at point $Q = \overline{OA} - \overline{OB_0}$. It is produced by changing the scan direction from the center to the left along the lower boundary of the scanning domain from point $A_0$ to point $B_0$, while the observation direction stays at point $A$ (zenith, $\alpha = 90°$). Segment II (shown as a blue line) starting at point $Q$ and ending at $M = \overline{OB} - \overline{OB_0}$ is produced by changing the observation direction from $A$ to $B$. Point $B$ is defined by the condition $|\partial k_y/\partial k_z|_B = |\partial k_{y_0}/\partial k_{z_0}|_{B_0}$ and has to be found numerically. Finally, Segment III (shown as a thick red line) starts at point $M$ and ends at $S = \overline{OC} - \overline{OC_0}$. This segment is computed by simultaneously changing the observation and scanning directions while preserving the equality between $|\partial k_y/\partial k_z|$ (on arc BC) and $|\partial k_{y_0}/\partial k_{z_0}|$ (on curve $B_0C_0$).

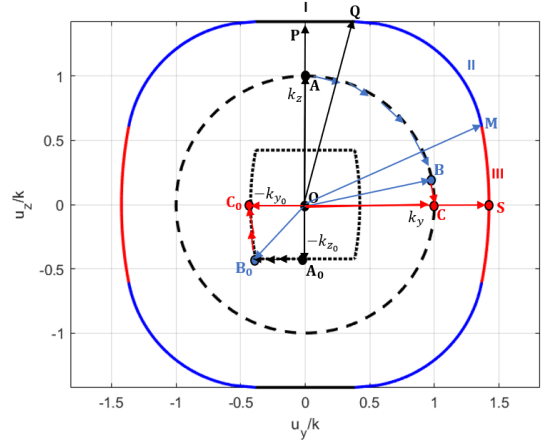

Fig. 3. Boundaries of the normalized $\boldsymbol{u}$-plane support. Black dotted line represents the boundary of the scan sector ($\boldsymbol{k}_0$), black dashed line circle depicts the observation domain ($\boldsymbol{k}$), and the solid lines represents $\boldsymbol{u}$-plane support.

Once the outer boundaries of the domain of interest in the $\boldsymbol{u}_T$-plane have been determined, the AF needs to be evaluated over the whole domain with a density sufficient to fully describe all sidelobes. Based on the Nyquist criterion, the sampling rates vs. $u_y$ and $u_z$ coordinates should be $\Omega L/2\pi$ and $\Omega H/2\pi$ where $L$ and $H$ denote the length and the height of the antenna, respectively [35]. Here, $\Omega > 1$ is the oversampling ratio selected to insure accurate interpolation. For example, in our computations we used $\Omega = 4$. The total number of sampling points is then given by $\Omega^2 LHA_u/4\pi^2$ where $A_u$ is the area of the domain of interest in the $\boldsymbol{u}$-plane.

The AF computation can be further accelerated using fast Fourier transform (FFT) for arrays with regular periodic lattices. More general cases such as aperiodic lattices can be tackled using a multilevel computational scheme proposed in [36]. Efficient AF characterization approach developed in this section facilitates the application of beam broadening techniques described below.

## IV. Beam Broadening by Phase Tapering Optimization

This section presents an example of phase tapering optimization for beam broadening providing a smooth phase distribution by optimization of only a small number of control parameters. In general, optimizing the excitation coefficients' phases is challenging since the problem is non-convex, while an extensive search over the space of all possibilities is impractical. The most popular optimization approach to the beam broadening problem found in the literature is to use various stochastic algorithms such as the GAs, particle swarm, and stochastic gradient descent. Although stochastic algorithms are practical for non-convex problems, they have some disadvantages. They can produce unstable results sensitive to errors and at times converge to local minima. Also, they consume significant computation time and memory which strongly depend on the number of elements.

Here, we present a simple and efficient polynomial phase tapering optimization (PTO) whose complexity depends mainly on the number of polynomial/coefficients and only weakly on the array's size. This method resembles that in [33], although extending the space to all even polynomials, while using a different cost function. The beam broadening is achieved by phase tapering introduced via excitation coefficients:

$$I_n(y, z) = e^{-j[P_y(y_n) + P_z(z_n)]} \quad (3)$$

which for simplicity, are assumed to have separable phase dependences along the $y$ and $z$ coordinates. Here, $P_\zeta(\zeta_n) = \sum_{i=1}^{N_\zeta^p} p_{\zeta,i} \zeta_n^{2i}$, $\zeta = y, z$, with $N_\zeta^p$ being the number of terms and $p_{\zeta,i}$ denoting the $i$th coefficient, are polynomials comprising only even powers to produce symmetric phase distributions. We optimize the polynomial coefficients to minimize the cost function:

$$f_{\text{cost}} = -\frac{P_{\text{ML}}}{P_{\text{SL}}} + c_1(BW_\varphi - BW_{\varphi_d})^{2a_1}$$
$$+ c_2(BW_\alpha - BW_{\alpha_d})^{2a_1} + c_3 R \quad (4)$$

which comprises four terms: the ratio between the energy in the ML to that in the sidelobe (SL) region, the deviation of the obtained azimuth and elevation beamwidths from the desired ones, and the maximum ripple in the ML. In (4), $P_s = \iint_{\boldsymbol{u} \in S} |\text{AF}(u_y, u_z)|^2 du_y du_z$, where $S = \text{ML, SL}$ designates the region in the $\boldsymbol{u}$-plane, $R$ refers to the maximum ripple, while $BW_\varphi/BW_\alpha$ and $BW_{\varphi_d}/BW_{\alpha_d}$ denote the computed and desirable azimuth/elevation beamwidths, respectively. Also, coefficients $c_i$ are positive real numbers and powers $a_i$ are positive integers.

Applying the PTO method, we only need to optimize $N_\zeta^p$ polynomial coefficients ($p_{\zeta,i}$). In addition, since PTO provides a continuous phase distribution, it is expected to be less sensitive to mutual coupling and defective elements than the results of stochastic optimization algorithms.

## V. Beam Broadening by Time-varying Phase Tapering

This section presents the TPT algorithm, which effectively produces amplitude tapering by time variation of the AF and combining of received returns from different pulses. This equivalence of the time varying phase tapering and conventional amplitude tapering is significant since it implies that pattern synthesis, such as beam broadening, can be achieved using an analytic solution well-known for the amplitude tapering.

Consider a pulse train with a pulse repetition interval (PRI) $T$ comprising an even number of pulses, $N_p$, and let $I_n^o$ and $I_n^e$ be the excitation coefficients of the odd and even pulses, respectively. Here, we set $I_n^o = \exp(-j\cos^{-1}(w_n))$ and $I_n^e$ be its complex conjugate, namely, $I_n^e = \exp(j\cos^{-1}(w_n))$, where $w_n$ is a desired amplitude weighting normalized such that $|w_n| \leq 1$. Thus, the transmission array factors of the odd/even pulses are given by:

$$\text{AF}_T^{o,e} = \sum_{n=1}^{N_e} e^{\mp j\cos^{-1}(w_n)} e^{j(u_y y_n + u_z z_n)} \quad (5)$$

where $\mp$ signs correspond to the odd/even pulses, respectively. Summing the target returns produced by odd and even pulses upon reception while using a time-independent receiving pattern, $\text{AF}_R$, and assuming stationary targets, we essentially add up odd and even AFs, thus getting an effective transmission AF:

$$\text{AF}_T^{\text{eff}} = \frac{1}{2}(\text{AF}_T^o + \text{AF}_T^e) = \sum_{n=1}^{N_e} w_n e^{j(y_n u_y + z_n u_z)} \quad (6)$$

One can see in (6) that an effective amplitude tapering was obtained by using only time dependent phase control.

So far, we have provided a method that produces the desired amplitude tapering by phase modulating the excitation coefficients and combining the received returns from stationary targets. Now, we will generalize the technique for a radar target with a relative radial velocity of $v_r$. In conventional pulse-Doppler radars, coherent integration is performed assuming that the velocity and radar cross section (RCS) are constant during the TOT. Under such assumptions, the received signal from a target at an initial range of $R_0$ is given by:

$$s(t) \propto \text{AF}_T \text{AF}_R e^{-j2k(R_0 - v_r t)} \quad (7)$$

In a typical case where $\text{AF}_T$, $\text{AF}_R$ are constant during the TOT, the only difference between adjacent pulses is the Doppler shift, i.e., $s(t + T) = s(t) e^{j2k v_r T}$. Then, FFT is performed on a sequence of $N_p$ sampled pulses for coherent integration [37-39]. In our method, we suggest modulating the phases of the excitation coefficients in the transmitting mode, meaning the

AF_T does not remain constant vs. time, precluding to use of FFT for coherent integration as usual. To implement coherent integration in conjunction with the TPT, we suggest a simple modification in the radar processing. As proposed above, we alternate the transmitting mode AF as given by (5). Then, we perform FFT on odd and even pulses separately to obtain:

$$S^o(l) \propto AF_T^o \sum_{n=0}^{\frac{N_p}{2}-1} \exp\left(jn\left(4kvT - \frac{4\pi}{N_p}l\right)\right) \quad (8a)$$

and

$$S^e(l) \propto AF_T^e e^{j2kv_rT} \sum_{n=0}^{\frac{N_p}{2}-1} \exp\left(jn\left(4kvT - \frac{4\pi}{N_p}l\right)\right) \quad (8b)$$

where $l$ is the frequency bin index. Finally, we sum the FFTs of the odd and even pulse sequences together with a proper compensation phase to obtain both coherent integration and the effective transmission AF of (6):

$$S^{eff}(l) = S^o(l) + e^{-j2kv_rT}S^e(l) \propto AF_T^{eff} \quad (9)$$

The compensation phase stems from the Doppler phase shift between adjacent pulses due to the target velocity $v_r$. The radial target velocity can be either roughly calculated based on the bin index as $v_r = l\pi/kN_pT$ or more accurately estimated using one of the frequency estimation techniques such as the pencil of functions [40].

Here, we successfully demonstrated equivalence between the time-varying phase and amplitude tapering while achieving coherent integration for moving targets, thus implying that the TPT algorithm is well suited for radar signal processing. The disadvantage of proposed algorithm is that the maximum unambiguous velocity is reduced by a factor of two since the effective PRI is now $2T$ [37-39].

## VI. NUMERICAL RESULTS

In this section, we apply the beam broadening techniques to a planar $16 \times 16$ element array with inter-element spacings of $\Delta_y = \Delta_z = 0.5\,\lambda$. While the beam broadening can be performed in both azimuth and elevation planes, for the sake of simplicity, here we apply it only in the elevation. Also, in the examples below we assume that the elevation and azimuth scan sector is defined by $\alpha_{0\min} = -25°, \alpha_{0\max} = 25°$, and $\varphi_{0\min} = -25°, \varphi_{0\max} = 25°$, respectively. We provide three different beam broadening levels: narrow, moderate, and wide using both the PTO and TPT techniques.

As a preliminary step, we reduce the dimensionality of the problem from five variables ($k, \alpha, \alpha_0, \varphi, \varphi_0$) to two - $u_y$ and $u_z$ and determine the relevant domain in the $\boldsymbol{u}$-plane as described in Section III. This is crucial for the optimization and helpful in displaying the results for all scan directions on a single plot.

Subsequently, we implement the PTO algorithm. As indicated in (3), the phase tapering that is exclusively oriented along the z-direction can be described by a four-term polynomial ($N_z^p = 4$). This yields excitation coefficients $I_n = e^{-j(\sum_{i=1}^{4} p_{z,i} z_n^{2i})}$. For the elevation only broadening, the cost function (4) reduces to $f_{cost} = -P_{ML}/P_{SL} + c_1(BW_\alpha - BW_{\alpha_d})^{2a_1} + c_2 R$. The choice of coefficients $c_1$ and $c_2$ entails striking a balance between attaining the desired beamwidth, minimizing undesired ripple effects, and maximizing the energy within the main lobe relative to the sidelobe region. We optimized the coefficients $p_{z,i}, i = 1, \ldots, N_z^p$ for $BW_{\alpha_d} = 10°, 30°, 50°$. Here, we used FMINSEARCH from the MATLAB optimization toolbox, an algorithm that finds the minimum of a function with a simplex search method [41]. Since the problem is not convex, we initialized the search for the polynomial coefficients randomly multiple times and set the optimal coefficients to be those that minimize the cost function. Fig. 4 illustrates the outcomes obtained using the PTO algorithm, while Fig. 5 showcases an instance of phase distribution described by a 4-term polynomial for the narrow beam broadening case.

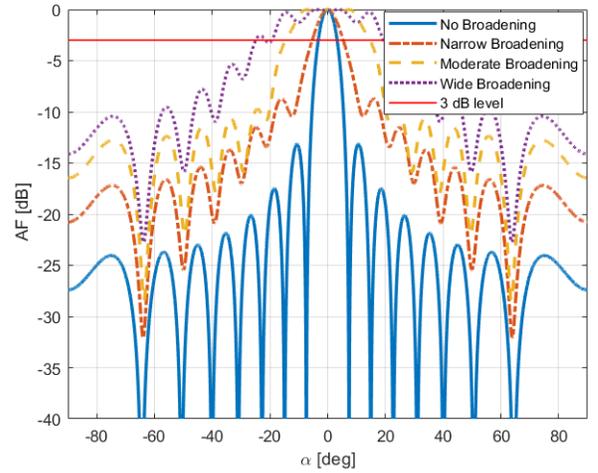

Fig. 4. Radiation patterns in the elevation cut for various levels of beam broadening using PTO.

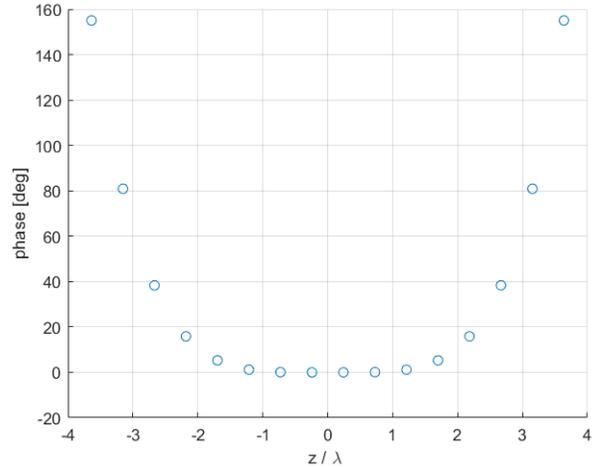

Fig. 5. Phase dependence on the $z$ coordinate for beam broadening obtained by the PTO technique for $BW_{\alpha_d} = 10°$.

In the PTO approach, the desired beam broadening was obtained relatively easily by determining four parameters, employing a simple and efficient optimization technique. However, it is noteworthy that the SLLs are quite high, especially for the wide broadening case. Figure 6 illustrates the AF over the $\boldsymbol{u}$-plane support, for the case of narrow broadening scenario. This serves as an illustrative example, as the scan sector's significance is reduced in other cases. The rationale behind plotting in the $\boldsymbol{u}$-plane is to visually demonstrate, in a single plot, the absence of grating or otherwise high side lobes for all scan angles. Indeed, Fig. 6 provides the evidence that no grating lobes arise in any scan direction.

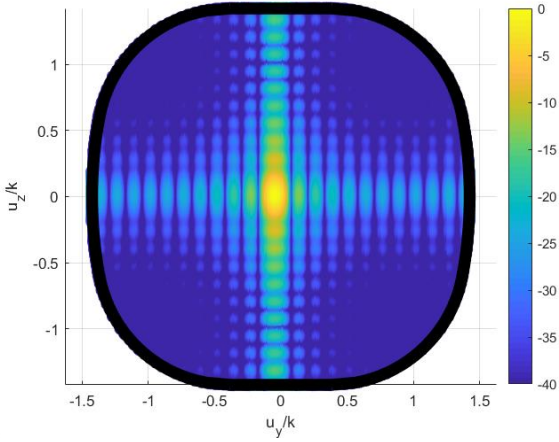

Fig. 6. Radiation pattern in the $\boldsymbol{u}$-plane obtained using PTO for the case of narrow beam broadening.

Next, we present the results for the TPT algorithm. In line with Section V, we employ a time varying phase tapering to produce effective amplitude tapering. Here, we consider a moving target with $kv_rT = 0.1134\pi$ and $N_p = 256$, for narrow, moderate, and wide beam broadening. We demonstrate the synthesis of a beam broadened pattern using a simple analytical solution. To further simplify the procedure, we consider a separable tapering where the excitation coefficient can be expressed as a product of separate $y$ and $z$ tapering functions, $I_{mn} = I_m^y I_n^z$. In this context, we assume a rectangular shape for the antenna array, where the total number of elements, $N_e$, is equal to the product of the number of elements along the $y$-axis, $N_y$, and the number of elements along the $z$-axis, $N_z$. Thus, rewriting the array factor expression:

$$\text{AF} = \sum_{m=1}^{N_y} I_m^y e^{jy_m u_y} \sum_{n=1}^{N_z} I_n^z e^{jz_n u_z}$$
$$= \text{AF}^y(u_y)\text{AF}^z(u_z) \quad (10)$$

where $\text{AF}^\zeta(u_\zeta), \zeta = y,z$ designate the one-dimensional array factors. The ideal desired $\text{AF}_D^\zeta(u_\zeta)$ pattern for the beam broadening problem is:

$$\text{AF}_D^\zeta(u_\zeta) = \begin{cases} 1 & |u_\zeta| \leq u_d^\zeta \\ 0 & |u_\zeta| \geq u_d^\zeta \end{cases} \quad (11)$$

where $u_d^y = k\sin\varphi_d$ and $u_d^z = k\sin\alpha_d$ are the desired beamwidths for broadening in azimuth and elevation, respectively. The excitation coefficient denoted as $I_n^\zeta$ can be obtained through the inverse discrete Fourier transform (IDFT) of $\text{AF}_D^\zeta$, resulting in the Dirichlet Kernel function (DK), namely $I_n^\zeta = \text{DK}(u_d^\zeta \zeta_n/2\pi)$. The discontinuities present in expression (11), specifically $\text{AF}_D^\zeta$ at points $\pm u_d^\zeta$, results in high SLL and ripples. A more effective strategy would be to taper the excitation coefficients accordingly:

$$I(\zeta_n) = \text{DK}\left(\frac{u_d^\zeta \zeta_n}{2\pi}\right) W(\zeta_n) \quad (12)$$

where $W(\zeta_n)$ is a SLL reduction windows function. The Hamming window has been selected as our choice, which serves as a suitable compromise between low SLL and better precision $\left(BW_{\alpha_d} \approx 2\alpha_d, BW_{\varphi_d} \approx 2\varphi_d\right)$. In Fig. 7, the results are illustrated for the specified $\alpha_d = 10°, 15°, 25°$. These angles approximately correspond to obtaining beamwidths $(BW_\alpha)$ of 20°, 30°, and 50° along the elevation cut. One can see from Fig. 7 that the ripples and the SLL in the TPT implementation are significantly lower than those obtained by using the PTO.

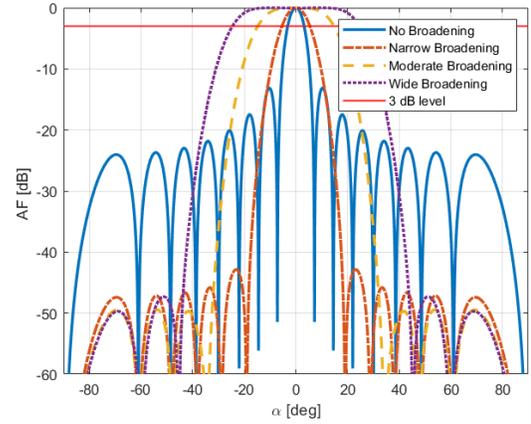

Fig. 7. Radiation patterns in elevation cut for various levels of beam broadening using TPT.

The phase variation of the excitation coefficients $I_n^o$ and $I_n^e$ along the $z$-axis for the wide broadening case is depicted in Fig. 8. This serves as an illustrative example, as the wide broadening case demonstrates the lowest effective antenna size compared to other scenarios. It is interesting to note that, even though the effective amplitude of many excitation coefficients approaches zero, the antenna is still transmitting at maximum power.

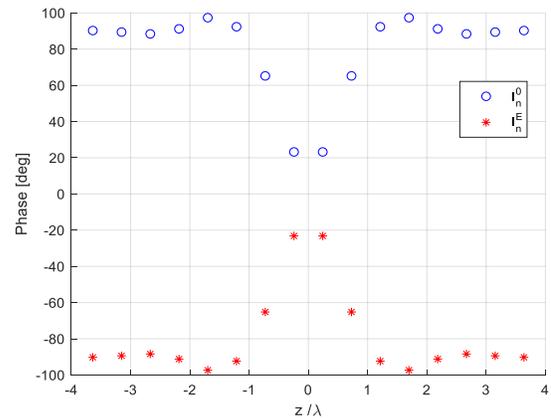

Fig. 8. The phase variation of excitation coefficients, $I_n^o$ and $I_n^e$.

As we have seen, the TPT method provided better results in terms of SLL and ripples as compared to the PTO, i.e., a time independent reference solution obtained through optimization. In addition, the TPT is simple to design mathematically, at least, in a separable case with a closed-form analytic solution.

## VII. CONCLUSION

This paper introduces an efficient AF evaluation method for AESA problems, along with two methods for beam broadening that exclusively employ phase tapering. The proposed TPT algorithm offers an efficient approach for implementing amplitude tapering using only phase shifters, providing pattern synthesis designs with very low SLL by a relatively simple adjustment in radar processing. We anticipate that this approach will be beneficial for beam shaping techniques in many applications.

## APPENDIX A

This appendix presents a detailed procedure for determining the boundary of the domain of interest in the $u_y u_z$ plane, where $u_y = k_y - k_{y0}$ and $u_z = k_z - k_{z0}$. Without loss of generality, we analyze the problem for positive values of $u_y$ and $u_z$. We will divide the solution into three segments, as depicted in Fig. 3.

Segment I corresponds to scanning of the main beam in azimuth over the range $\varphi_{0_{min}} \leq \varphi_0 < 0°$ for $\alpha_0 = \alpha_{0_{min}}$ while keeping the observation direction fixed at zenith $k_z = k$ ($\alpha = 90°$) and $k_y = 0$ (point A). Here, $k_{z_0} = k \sin(\alpha_{0_{min}})$ remains constant, while the absolute value of $k_{y_0} = k \cos(\alpha_{0_{min}}) \sin(\varphi_0)$ increases from 0 to $k_{y_0} = k \cos(\alpha_{0_{min}}) \sin(\varphi_{0_{min}})$ (segment $A_0 B_0$). Thus, the $yz$ components of the difference vector $\boldsymbol{u}_T$ are given by $u_y = k \cos(\alpha_{0_{min}}) \sin(\varphi_0)$ and constant $u_z = k(1 + \sin|\alpha_{0_{min}}|)$.

Segment II corresponds to the variation of the observation direction from $A$ to $B$ along the arc $AB$ while keeping the main beam direction fixed at point $B_0$. Along this path, the derivative $|\partial k_y / \partial k_z|$ is decreasing until it is equal to $|\partial k_{y_0} / \partial k_{z_0}|$. We express $k_y$ and $k_{y_0}$ in terms of $k_z$ and $k_{z_0}$, respectively, i.e., $k_y = \sqrt{k^2 - k_z^2}$ and $k_{y_0} = \sqrt{k^2 - k_{z_0}^2} \sin(\varphi_0)$. The partial derivatives are:

$$\frac{\partial k_y}{\partial k_z} = -\frac{k_z}{\sqrt{k^2 - k_z^2}} = -\tan(\alpha) \quad (A.1a)$$

$$\frac{\partial k_{y_0}}{\partial k_{z_0}} = -\frac{k_{z_0} \sin(|\varphi_0|)}{\sqrt{k^2 - k_{z_0}^2}} = -\tan|\alpha_0| \sin|\varphi_0| \quad (A.1b)$$

At point $B_0$, the value of the partial derivative $|\partial k_{y_0} / \partial k_{z_0}|$ is obtained for $\alpha_0 = \alpha_{0_{min}}$ and $\varphi_0 = \varphi_{0_{min}}$. Segment II ends at point $M$ when the partial derivatives are equal:

$$\left|\frac{\partial k_y}{\partial k_z}\right| = \left|\frac{\partial k_{y_0}}{\partial k_{z_0}}\right|\bigg|_{\alpha_0 = \alpha_{0_{min}}, \varphi_0 = \varphi_{0_{min}}} \quad (A.2)$$

and, consequently, point $B$ is defined by:

$$\alpha_B = \tan^{-1}(\tan|\alpha_{0_{min}}| \sin|\varphi_{0_{min}}|) \quad (A.3)$$

where $\alpha_B$ is the corresponding elevation angle. For Segment II, three parameters are held constant: $\alpha_0$ is set to $\alpha_0 = \alpha_{0_{min}}$, $\varphi_0$ is set to $\varphi_0 = \varphi_{0_{min}}$, and $\varphi$ is fixed at 90°. Additionally, $\alpha$ starts at 90° and decreases to $\alpha_B$.

Over Segment III – arc MS, we change $k_z$ and $k_{z_0}$ simultaneously along arcs $BC$ and $B_0 C_0$, respectively, while keeping the derivatives $|\partial k_y / \partial k_z|$ and $|\partial k_{y_0} / \partial k_{z_0}|$ equal at every step. Practically, we increase the scanning elevation $\alpha_0$ by small steps from $\alpha_{0_{min}}$ to 0 and find the corresponding observation elevation angle:

$$\alpha = \tan^{-1}(\tan|\alpha_0| \sin|\varphi_{0_{min}}|) \quad (A.4)$$

Thus, for Segment III, $\varphi = 90°$, $\varphi_0 = \varphi_{0_{min}}$, $\alpha_{0_{min}} \leq \alpha_0 \leq 0$, and $\alpha$ satisfies (A.4).